\documentclass[aps]{revtex4}
\def\bc{\begin{center}}
\def\ec{\end{center}}

\def\beq{\begin{equation}}
\def\eeq{\end{equation}}

\def\br{{\bf r}}

\begin{document}


\title{Non-Abelian chiral symmetry controls random scattering in two-band models\\
}

\author{K. Ziegler}

\affiliation{
Institut f\"ur Physik, Universit\"at Augsburg\\
D-86135 Augsburg, Germany
}
\date{\today}

\begin{abstract}
We study the dynamics of non-interacting quantum particles with two bands in the presence 
of random scattering. The two bands are associated with a chiral symmetry. After breaking the latter
by a potential, we still find that the quantum dynamics is controlled by a non-Abelian chiral 
symmetry. The possibility of spontaneous symmetry breaking is analyzed within a self-consistent
approach, and the instability of a symmetric solution is discussed.
\end{abstract}

\maketitle

\section{Introduction}

Recent experiments on two-dimensional two-band systems, such as graphene \cite{novoselov05,zhang05}
or the surface of topological insulators \cite{bouvier11,he11}, 
have revealed that transport in these systems is very robust with a universal behavior that does not 
depend on the sample quality. 
In general,
the mechanism for the appearance of universal and robust properties in many-body systems is often 
associated with underlying symmetries and the spontaneous breaking of these symmetries. For instance,
the long-range and long-time behavior is usually controlled by symmetries but it is not 
much affected by the fine details of the model. Apart from critical points in systems with 
second-order phase transitions, dynamical properties such as diffusion may also be related to 
symmetry properties. An example is a system of non-interacting Dirac fermions in two dimensions at 
the Dirac node with a random gap, where we have identified a non-Abelian chiral symmetry 
\cite{ziegler98a}. This symmetry transformation is usually called ``supersymmetric'' because it
connects bosons and fermions \cite{efetov96}. It turned out that spontaneous breaking of this symmetry generates
a massless fermion mode, caused by pairing a boson and a fermion. This is analogous to
the Cooper pair in superconductivity which consists of a pair of fermions rather than a 
boson-fermion pair. Another difference is that the massless fermion decays as a power law and 
does not provide a coherent state. The physical effect of the massless mode is diffusion of fermionic 
quasiparticles. On the other hand, the system is insulating if the symmetry is not spontaneously broken.
This can happen for a sufficiently large average gap \cite{Ziegler1997}. Moreover, diffusion and the 
metal-insulator transition are described by a simple scaling law which deviate from the 
Drude behavior of conventional metals \cite{Ziegler2009}.
All this indicates that the spontaneous breaking of the non-Abelian chiral symmetry is the origin of the
metallic behavior and the metal-insulator transition due to random gap opening of Dirac fermions. 

The idea of this work is to extend the analysis of the Dirac fermions to the more general case
of tight-binding models with two bands. In particular, we will identify the non-Abelian chiral symmetry
for a four-body Hamiltonian which is constructed from one-body Hamiltonians with broken chiral 
symmetry.

\section{Model}

In a chiral-invariant system states in the upper band at energy $E$ are directly connected with states in the lower 
band at energy $-E$ by a linear transformation $U$. In other words, the chiral 
Hamiltonian satisfies the anti-commutator relation
\beq
[U,H]_+=0
\ .
\label{chiral}
\eeq
For an eigenstate $\Psi_E$ of energy $E$ with $H\Psi_E=E\Psi_E$ this relation implies that $\Psi_{-E}=U\Psi_E$;
i.e., the states of the negative (positive) spectrum are created by the linear transformation $U$ from the states
of the positive (negative) spectrum. A consequence of the vanishing anti-commutator is the chiral symmetry relation
\beq
e^{\alpha U}He^{\alpha U}=H
\label{cont_chiral}
\eeq
with a continuous parameter $\alpha$.

\subsection{Special Hamiltonians}

A two-band Hamiltonian
can be represented in terms of Pauli matrices $\sigma_j$ ($j=0,...,3$) as
\beq
H=h_0\sigma_0+h_1\sigma_1+h_2\sigma_2+h_3\sigma_3
\ ,
\label{hamilton00}
\eeq
where $h_j$ are matrices in real space, and the associated eigenfunctions of $H$ are spinor states.
A symmetric example is a tight-binding Hamiltonian on a bipartite lattice, such as the honeycomb lattice.
It can also be written in the form of Eq. (\ref{hamilton00}), where the Pauli matrices refer to the 
two sublattices. For nearest-neighbor hopping on the honeycomb lattice the Fourier components of the 
coefficients read
\[
h_1=-t\sum_{j=1}^3\cos(a_j\cdot k), \ \ \ h_2=-t\sum_{j=1}^3\sin(a_j\cdot k)
\]
with the lattice vectors $a_1=(-\sqrt{3}/2,1/2)$, $a_2=(0,-1)$ and $a_3=(\sqrt{3}/2,1/2)$. $h_3$ either plays the role of a
staggered potential that breaks the sublattice symmetry, or it represents a next nearest-neighbor hopping term
on the honeycomb lattice \cite{kadirko11}.
For $h_3=0$ the Hamiltonian has a chiral symmetry with $U=\sigma_3$ while for $h_3\ne0$ the chiral symmetry is broken.
We will treat Hamiltonians with broken chiral symmetry in Sect. \ref{sect:symmetrybreaking}.

{\it Remark}: A prominent example for Eq. (\ref{hamilton00})
is the two-dimensional Dirac Hamiltonian, where $h_j=i\partial_j$ ($j=1,2$) and $h_3=m$ is the mass term.
It should be noticed that the Dirac Hamiltonian is not symmetric but chiral invariant
for $m=0$, and it satisfies Eq. (\ref{chiral}) for $U=\sigma_3$ again. $m\ne 0$ breaks the chiral
symmetry. In that case the Dirac Hamiltonian satisfies 
\beq
\sigma_1 H=-H^*\sigma_1
\ ,
\label{broken_chiral}
\eeq
rather than the relation (\ref{chiral}), with $H^*=H^T$, where $^T$ with the matrix transposition \cite{ziegler98a}. 
This relation is associated with particle-hole symmetry \cite{zirnbauer96}.

\section{few-body approach
}

Now we consider briefly the dynamics of non-interacting particles in a random environment and derive a 
four-body Hamiltonian to describe transport behavior in the presence of random scattering.
In order to study the motion of a quantum particle in space we calculate the transition
probability of a particle going from site $\br$ at time $t$ to site $\br'$ at time $t'$. 
The transition amplitude (or Green's function) for this process is
\[
G_{\br,t;\br',t'}=\int e^{-iE(t-t')}(H-E+i\epsilon)^{-1}_{\br,\br'}dE
\ ,
\]
and the corresponding transition probability reads 
\beq
P_{\br,t;\br',t'}=|G_{\br,t;\br',t'}|^2
=\int e^{-i(E-E')(t-t')}(H-E+i\epsilon)^{-1}_{\br,\br'}(H-E'-i\epsilon)^{-1}_{\br',\br}dEdE'
\ .
\label{2pgf}
\eeq
Then the expansion of the wave function within the time interval $[t,0]$ can be written as
\[
\langle r_j^2\rangle=\frac{\sum_\br r_j^2 P_{\br,t;0,0}}{\sum_\br P_{\br,t;0,0}}
\ .
\]
A straightforward calculation for a translational invariant Hamiltonian $H$ results in a quadratic
increase of this expression with time $t$. This is ballistic transport, in contrast to diffusion,
where it would increase linearly with time. Diffusive transport is possible when we consider a random
Hamiltonian $H$ and average the transition probability with respect to the distribution of $H$.

The product of the two resolvents in Eq. (\ref{2pgf}) can be interpreted as a scattering process 
of two independent particles that move in opposite time directions because of $\pm i\epsilon$
(advanced and retarded Green's functions). As a result, the dynamics is described by the 
scattering of two independent particles with identical Hamiltonians. In principle, we could 
add more independent particles, since we always can project
the resulting few-body Hamiltonian to the desired product of Green's functions. Using a functional-integral
representation of non-interacting particles (cf. Appendix \ref{sect:app_1}), we can 
accommodate an arbitrary number of factors.
It is important to notice that the average of the Green's function leads to an effective interaction between
the particles. We assume that the one-body Hamiltonians obey the relation (\ref{chiral}). 
Now we can take advantage of the relation (\ref{chiral}) 
which allows us to switch between
retarded and advanced Green's functions by applying the linear transformation $U$. This is discussed
in the next Section.

\subsection{non-Abelian chiral symmetry of the four-body Hamiltonian}
\label{sect:symmetrybreaking}

We consider the extended Hamiltonian $H+H_1$, which is not chiral symmetric, and the transformation matrices 
$U_j$ ($j=1,2$), for which we assume that they satisfy the (anti-)commutator relation
\beq
[H,U_j]_+=[H_1,U_j]_-=0
\label{relations1}
\eeq
as a generalization of the relation (\ref{chiral}). This includes the case $U_1=U_2\equiv U$.
Next we replicate the one-body Hamiltonian $H+H_1$ to a two-body Hamiltonian ${\bar H}+{\bar H}_1$ with
\[
{\bar H}=\pmatrix{
H & 0 \cr
0 & H \cr
}, \ \ \ 
{\bar H_1}=\pmatrix{
H_1 & 0 \cr
0 & -H_1 \cr
}, \ \ \ 
{\bar U}=\pmatrix{
U_1 & 0 \cr
0 & U_2 \cr
}
\ .
\]
Together with Eq. (\ref{relations1}) these expressions are connected by the relations
\beq
[{\bar H},{\bar U}]_+=[{\bar H_1},{\bar U}]_-=0
\ .
\label{relations2}
\eeq
Now we introduce the four-body Hamiltonian
\beq
{\hat H}=\pmatrix{
{\bar H}+{\bar H_1} & 0 \cr
0 & {\bar H}-{\bar H_1} \cr
}
\label{ham_structure}
\eeq
and the transformation matrix
\beq
{\hat U}=\pmatrix{
0 & 0 & \varphi_1U_1 & 0 \cr
0 & 0 & 0 & \varphi_2U_2\cr
\varphi_1'U_1 & 0 & 0 & 0 \cr
0 & \varphi_2'U_2 & 0 & 0 \cr
}
\label{chiral1}
\eeq
and obtain the anti-commutator relation
\beq
[{\hat H},{\hat U}]_+=0
\ .
\label{relations3}
\eeq
This implies the non-Abelian chiral symmetry
\beq
e^{\hat U}{\hat H}e^{\hat U}={\hat H}
\ .
\label{na_chiral}
\eeq
Moreover, we notice that $\det({\bar H}+{\bar H_1}+i\epsilon)=\det({\bar H}-{\bar H_1}+i\epsilon)$.
Then we obtain for the graded determinant \cite{Ziegler2009} 
\[
{\rm detg}({\hat H}+i\epsilon)= {\rm detg}(e^{\hat U})=1
\ .
\]
Eventually, there is a relation between advanced and retarded Green's function:
\beq
{\hat U}({\hat H}+i\epsilon)^{-1}{\hat U}^{-1}=(-{\hat H}+i\epsilon)^{-1}
=-({\hat H}-i\epsilon)^{-1}
\ .
\eeq
Eqs. (\ref{relations3}) and (\ref{na_chiral}) are a non-Abelian generalization of Eqs. (\ref{chiral}) 
and (\ref{cont_chiral}), respectively.
The symmetry transformation in Eq. (\ref{chiral1}) depends on the free parameters $\varphi_j,\varphi_j'$.
For the special case of a random Hamiltonian they are Grassmann variables (cf. Appendix \ref{sect:app_1} and 
Refs. \cite{ziegler98a,Ziegler2009,ziegler12c}). This can be understood as if the upper block matrix
(the lower block matrix) 
of the Hamiltonian (\ref{ham_structure}) acts on bosons (fermions). Consequently, the off-diagonal blocks
in the transformation matrix ${\hat U}$ transform bosons into fermions and vice versa. This requires that
the parameters $\varphi_j,\varphi_j'$ are Grassmann variables.

{\it Remark:}
The construction of the Hamiltonian ${\hat H}$ would work for $H=0$. This implies that we could obtain
the non-Abelian chiral symmetry even for a one-body Hamiltonian with only one band. The reason is that ${\bar H}$ 
is a two-body Hamiltonian with chiral symmetry.

\section{Spontaneous symmetry breaking
}

The main question is whether or not the non-Abelian chiral symmetry in Eq. (\ref{na_chiral}) is 
spontaneously broken for the average Green's function $\langle({\hat H}+i\epsilon)^{-1}\rangle$
after sending $\epsilon\to0$. This can be studied by employing the self-consistent Born approximation 
(SCBA), which is equivalent to the saddle-point approximation of the functional integral in (\ref{fint1}).
For this purpose we introduce the (complex) self-energy
$\eta$ to approximate the average Green's function 
by
\beq
\langle ({\hat H}+i\epsilon)^{-1}\rangle
\approx (\langle {\hat H}\rangle+i\epsilon+i\eta)^{-1}\equiv{\hat G}_0
\ .
\label{scba1}
\eeq
$\eta$ satisfies the self-consistent (or saddle-point) equation for disorder strength $g$
\beq
\eta=g\int_k\frac{z}{\lambda_1^2+\lambda_2^2+\lambda_3^2+z^2} ,\ \ \ z=\eta-i\mu
\ , 
\label{spe1}
\eeq
where $\lambda_j$ are the Fourier components of $h_j$ of the Hamiltonian. The integration is performed with
respect to the Brillouin zone.
$\mu$ is a uniform chemical potential which replaces the $H_1$ term in the Hamiltonian: $H_1\equiv\mu\sigma_0$.  
This equation is equivalent to
\beq
i\mu =z\left(-1+g\int_k\frac{1}{\lambda_1^2+\lambda_2^2+\lambda_3^2+z^2}\right)\equiv zf(z)
\ ,
\label{spe2}
\eeq
which allows us to determine the two solutions $z(\pm\mu)$ for a given $\mu$. 
The solutions of this equation can be distinguished according to their
transformation behavior under the sign change $\mu \to-\mu $: There is 
a symmetric solution with
\beq
z(-\mu )=-z(\mu )
\label{symm_sol}
\eeq
which does not break the non-Abelian chiral symmetry and an asymmetric solution that obeys
\beq
z(-\mu )\left(-1+g\int_k\frac{1}{\lambda_1^2+\lambda_2^2+\lambda_3^2+z(-\mu )^2}\right)
=-z(\mu )\left(-1+g\int_k\frac{1}{\lambda_1^2+\lambda_2^2+\lambda_3^2+z(\mu )^2}\right)
\ ,
\label{asymm_sol}
\eeq
where $z(-\mu )^2\ne z(\mu )^2$. The symmetric solution in (\ref{symm_sol})  implies $z(0)=0$.
Moreover, the effective symmetry breaking field is
\beq
{\bar\eta}=
\frac{z(\mu )+z(-\mu )}{2}=\frac{\eta(\mu )+\eta(-\mu )}{2}
\ .
\label{sb_field}
\eeq
Whether the symmetric or the asymmetric solution is relevant depends on the stability of the solution.
Inspection of the fluctuations around the symmetric solution, according to the expansion in Appendix \ref{sect:app_2},
provides us a stable symmetric solution if the kernel of the quadratic form in Eq. (\ref{stability}) is non-negative
and unstable symmetric solution if it has negative eigenvalues.
Thus spontaneous symmetry breaking is indicated by an instability of a symmetric solution. This will be discussed in the
next Section for a specific choice of the Hamiltonian $H$.

\subsection{Example: linear spectrum}

In the case of a linear spectrum with $s$ nodes, where we have $\lambda_j=k_j$ ($j=1,2$) with cut-off $\lambda$ 
(i.e. $0\le k_1^2+k_2^2\le\lambda^2$) and $\lambda_3=0$, the integral in Eq. (\ref{spe2}) can be performed to 
yield
\beq
\eta = z\frac{sg}{2}\log\left(1+\lambda^2/z^2\right)
\ .
\label{sceq1}
\eeq
For $\mu=0$ this equation has the symmetric solution $z=\eta=0$ and the asymmetric solution
\[
z=\eta=\frac{\lambda}{\sqrt{e^{2/sg}-1}}
\ .
\] 
For $z\ne0$ we can rewrite (\ref{sceq1}) as
\beq
\lambda^2=z^2\left( e^{2/sg}e^{2i\mu/sgz}-1\right)
\ .
\eeq
Thus, the imaginary part of the right-hand side must vanish and its real part must be $\lambda^2$.
There is no closed solution of these equations but for
$\mu\sim0$ we can expand the exponential function for the asymmetric solution and obtain a quadratic expression 
for the right-hand side 
\[
\eta=\left(1-\frac{\beta}{\alpha sg}\right)i\mu
\pm\sqrt{\frac{\lambda^2}{\alpha}+\mu^2\frac{\beta}{(sg)^2\alpha}(2-\beta/\alpha)}
\ \ \ (\alpha=\beta-1 ,\ \ \ \beta=e^{2/sg})
\ .
\]
Thus, the symmetry breaking field of (\ref{sb_field})
\[
{\bar\eta}
=\sqrt{\frac{\lambda^2}{\alpha}+\mu^2\frac{\beta}{(sg)^2\alpha}(2-\beta/\alpha)}
\]
increases with $\mu$. For small $g$ we have $\beta/\alpha\sim1$ such that
\beq
{\bar\eta}\sim
\sqrt{\frac{\lambda^2}{\alpha}+\frac{\mu^2}{(sg)^2}}
\ .
\eeq
In physical terms, ${\bar\eta}$ is the effective scattering rate in the Green's function (\ref{scba1}). 
Our result means that the scattering
rate increases as we go away from the nodes at $\mu=0$ into the bands. This effect is not surprising, since the
density of states also increases as we go away from the nodes. 
For this special case the eigenvalues of the stability matrix on large scales 
have been evaluated in Ref. \cite{Ziegler1997} as $(1/g,1/g+2,1/g-2,1/g)$. Thus, the symmetric solution 
is unstable for $g>1/2$.

\section{Summary and Conclusions}

The starting point of our analysis is a one-body Hamiltonian $H$ with chiral symmetry (\ref{cont_chiral}). 
After breaking this symmetry by a potential term, we have identified the non-Abelian chiral symmetry 
(\ref{na_chiral}) of the corresponding four-body Hamiltonian (\ref{ham_structure}). This four-body 
Hamiltonian can be used to describe random scattering processes, where the transformation from 
retarded to advanced Green's is performed by a chiral transformation.

The non-Abelian chiral symmetry can be spontaneously broken after averaging the Green's function with 
respect to random fluctuations of the Hamiltonian. We have studied the general case in a self-consistent 
approximation and have identified symmetric and asymmetric solutions. For a special Hamiltonian with
linear spectrum around the nodes we have found that the symmetric solution becomes unstable for sufficiently 
strong random fluctuations. 

These results indicate that the symmetric two-band Hamiltonian has a similar non-Abelian 
structure as the 2D Dirac fermions with particle-hole symmetry. Like for the latter case, we anticipate
that the spontaneously broken non-Abelian chiral symmetry leads to a massless fermion mode which describes
the physics on large scales.

\appendix

\section{Functional integral: distribution of the Green's function}
\label{sect:app_1}

Now we combine the matrix structure of ${\hat H}$ in Eq. (\ref{ham_structure}) with the averaging procedure 
for a random Hamiltonian $H$. This will justify that the parameters $\varphi_j,\varphi_j'$ in the transformation matrix
${\hat U}$ are Grassmann variables. For this purpose we employ a functional-integral approach with bosons and fermions. 
The latter has been discussed extensively in the literature \cite{efetov96} and specifically for the problem of two-band 
models in Refs. \cite{Ziegler1997,Ziegler2009,ziegler12c}. Therefore, only a brief summary that is relevant for the 
symmetry discussion, is given here whereas details can be found in the mentioned literature. 

We consider the Green's functions or products of Green's functions, averaged over a Gaussian distribution of some
physical quantity inside the Hamiltonian ${\hat H}$.
For simplicity we focus here on a random chemical potential and study $\langle ({\hat H}+i\epsilon)^{-1}\rangle$,
where we use the notation for the average
\beq
\langle ({\hat H}+i\epsilon)^{-1}_{r,r} \rangle 
=\frac{1}{{\cal N}}\int ({\hat H}+i\epsilon)^{-1}_{r,r} \prod_r e^{-\mu_r^2/g}d\mu_r
\label{average1}
\eeq
with the normalization ${\cal N}$.
Now we can apply a transformation from the random variable $\mu_r$ to the $8\times 8$ random matrix ${\hat Q}_r$
\beq
{\hat Q}_\br=\pmatrix{
Q_\br & \Theta_\br \cr
{\bar \Theta}_\br & iP_\br\cr
}
\label{supermatrix}
\eeq
with Hermitean $4\times4$ matrices $Q_\br$ and $P_\br$ and $4\times4$ supermatrices $\Theta$, ${\bar \Theta}$.
$Q_\br$ is related to the diagonal elements of the Green's function $({\hat H}+i\epsilon)^{-1}_{r,r}$ due to
\beq
\langle ({\hat H}+i\epsilon)^{-1}_{r,r}\rangle
=\int {\hat Q}_r J\prod_r e^{-Trg_4({\hat Q}_r^2)/g}{\cal D}_4[{\hat Q}_r]
\ .
\label{fint1}
\eeq
Products of Green's functions can also be represented by this functional integral \cite{Ziegler1997}.
$J$ is the Jacobian of the transformation $\mu_\br\to{\hat Q}_\br$ and reads as a graded determinant
\beq
J={\rm detg}(\langle{\hat H}\rangle+i\epsilon{\hat \tau}_0+{\hat Q}) \ \ \ 
\ .
\label{jacobian0}
\eeq
It is important to notice that $J$ is invariant under the global symmetry transformation (\ref{chiral1}) 
because ${\rm detg}(e^{\hat U})=e^{{\rm Trg}{\hat U}}=1$ with the graded trace ${\rm Trg}$. Then we can 
apply the transformation ${\hat Q}_\br\to e^{\hat U}{\hat Q}_\br e^{\hat U}$, which leaves the integral 
invariant in the limit $\epsilon\to0$. 
By sending $\epsilon$ to zero we are able to study spontaneous symmetry breaking
within a saddle-point approximation of the functional integral (\ref{fint1}).

\section{Instability of the symmetric saddle point}
\label{sect:app_2}

Fluctuations around the symmetric saddle point ${\hat Q}=0$ can become unstable when there is an 
asymmetric saddle point. Gaussian fluctuations around the saddle-point solution indicate clearly 
the instability of the symmetric solution \cite{Ziegler1997}.
The stability analysis can be applied to the integral in Eq. (\ref{fint1}) by expanding the logarithm of the 
Jacobian in Eq. (\ref{jacobian0}) around ${\hat Q}=0$:
\beq
\frac{1}{g}Trg_4({\hat Q}_r^2)-\log {\rm detg}(\langle{\hat H}\rangle+i\epsilon{\hat \tau}_0+{\hat Q})
= 
\frac{1}{g}Trg_4({\hat Q}_r^2)-\frac{1}{2}Trg_4\left[{\hat G}_{0;r,r'}{\hat Q}_{r'}{\hat G}_{0;r',r}{\hat Q}_{r}\right]
+o({\hat Q}^3)
\ .
\label{stability}
\eeq
The symmetric solution is stable if the kernel of the quadratic form in ${\hat Q}$ 
is non-negative (i.e. it has no negative eigenvalues). On the other hand, it is unstable if one or
more eigenvalues become negative.

\end{document}